\documentclass{article}
\usepackage{amsmath}
\usepackage{amsfonts}
\usepackage{geometry}
\usepackage{graphicx}
\usepackage{setspace} 

\begin{document}

\title{Benefits of Diversity, Communication Costs, and Public Opinion
Dynamics\thanks{The authors thank the National Bank of Belgium for financial
  support.}} 
\author{Gani Aldashev\thanks{Department of Economics and CRED, University of
    Namur (FUNDP). Email: gani.aldashev@fundp.ac.be.} \and Timoteo
Carletti\thanks{Corresponding author. Department of Mathematics, University of
  Namur 
(FUNDP). Telephone number: +3281724903, fax number: +3281724914. Mailing address: 8 Rempart de la Vierge,
5000 Namur, Belgium. Email: timoteo.carletti@fundp.ac.be.}}
\maketitle

\begin{abstract}
We study the dynamics of public opinion in a model in which agents change
their opinions as a result of random binary encounters if the opinion difference is
below their individual thresholds that evolve over time. We ground these
thresholds in a simple individual cost--benefit analysis with linear benefits
of diversity and quadratic communication costs. We clarify and deepen the
results of earlier continuous--opinion dynamics
models (Deffuant et al., Adv Complex Systems 2000~\cite{Deffuant}; Weisbuch et al., Complexity
2002~\cite{Weisbuch}) and establish
several new results regarding the patterns of opinions in the asymptotic
state and the cluster formation time.
\end{abstract}

\vspace{2em}{\it Keywords:} Opinion dynamics, agent--based models, diversity, communication
  costs. \\

\vspace{2em}{\it Number of text pages:} 14. {\it Number of Figures:} 7. {\it
  Number of Tables:} 2.
\newpage

\begin{doublespace}
\section{Introduction}

The study of the effect of public opinion on government policies has a long
tradition in social sciences. In a breakthrough article, Page and Shapiro~\cite{Page} find that
public opinion affects policies more than policies affect opinion. This effect
is particularly strong for large and sustained changes in public 
opinion and for policy issues that are more salient.

Given that public opinion is a major determinant of government policies, one
naturally asks: How public opinion is formed? What drives those large and
sustained changes in public opinion? In a recent study, Blinder and Krueger
~\cite{Blinder} find that ideology shapes public opinion more than
self--interest or economic knowledge. This implies that the
utility--maximizing rational--choice framework is perhaps not the most
adequate avenue for understanding public opinion. Indeed, discussing the
findings of Blinder and Krueger~\cite{Blinder}, Nordhaus hypothesizes 
that \lq\lq public opinions about the economy in a democracy are the outcome
of a 
very complex process in which people try to sort through conflicting
accounts and theories, often provided by unreliable
narrators\rq\rq~\cite{Nordhaus}. 

This statement indicates that the role of social
interactions is crucial in 
public opinion formation and dynamics. Following this line, a small but growing
interdisciplinary literature tries to model public opinion formation as a
process shaped by individual--level social interactions. Based on an early paper by Follmer~\cite{Follmer}, Orl\'{e}an~\cite{Orlean} and Ianni and
Corradi~\cite{Ianni} analyze the dynamics of public opinion assuming binary
opinion values (e.g. for or against a certain issue).
Orl\'{e}an~\cite{Orlean} finds that if social interactions are
non--sequential, public opinion dynamics is often non--ergodic, i.e. the
asymptotic behavior of the opinion formation process is not described by a
unique limit distribution, independent of the initial
conditions. Ianni and 
Corradi~\cite{Ianni} consider a model in which an individual tends to
conform with the opinion held by the majority of her neighbors. They study
several majority rules and describe under which rules public opinion
converges to a consensus (a single cluster of opinions) or becomes polarized (two
large clusters).

The insight provided by these papers is surely relevant for the understanding
of public opinion formation. However, they remain limited to binary
opinions, whereas most of real--world 
questions relate to multiple or continuous
opinions. For instance, while a citizen states that she is a liberal or a
conservative, the \textit{strength} of her position contains much richer
information about her preferences. On certain issues there is
probably more disagreement between a radical liberal and a moderate liberal
than between the latter and a moderate conservative. Moreover, on
certain political issues public opinion is divided between left, center, and
right (three clusters), which lies beyond the reach of the simple
binary--opinion models.

We are thus interested in understanding public opinion
formation under the assumption of continuous opinions. The literature in
this direction is scarce. Deffuant et al.~\cite{Deffuant}
and Weisbuch et al.~\cite{Weisbuch} study the dynamics of a model 
where agents adjust their opinions as a result of random encounters whenever
the difference in opinions is below a certain threshold. This threshold
captures the degree of openness of an agent towards opinions differing from
hers. They find that high thresholds lead to convergence towards consensus, while low thresholds generate several opinion
clusters.

So far, however, our understanding of public opinion formation is limited to this fundamental insight. The following crucial questions remain
open: Why in some contexts citizens are more open towards differing opinions
than in others? Does this openness vary between moderate citizens and their
more extremist counterparts? And if so, how the differences in relative
openness within a community affect public opinion dynamics?

This paper provides answers to these questions by adapting the model
of Deffuant et al.~\cite{Deffuant} in a simple but flexible way capturing various forces that may affect the relative openness of moderates and
extremists. Our basic focus is on the technological characteristics of
social interactions, as described by the benefits of diversity versus communication costs. This tension represents the fundamental trade--off of
openness. On one hand, being more open towards other opinions can bring in informational benefits that come from opinions built using very
different sources. On the other hand, being more open implies the obligation
to communicate with very different individuals, and in such case communication costs can be high. In this paper, we show how this
trade--off affects the patterns of public opinion and the dynamics of opinion formation process.

We also analyze the time needed for the population to
form the main opinion group in the consensus case and the two largest
groups in the polarization case. The cluster formation time is detemined, on
one hand, by the technological characteristics of interactions and, on the
other hand, by the size of community. In the case of polarized communities,
the fundamental openness trade-off plays the key role in the dynamics of opinion formation process.

\section{The model}

Consider a community populated by $N$ individuals. Each citizen holds at time
$t$ an opinion which is represented by a continuous
variable ranging between $0$ and $1$, $x_{i}^{t}\in \lbrack 0,1]$ for
$i=1,...,N$. Time evolves in discrete steps, and social interactions occur
randomly according to the following rule. In each time step, two
citizens are randomly drawn from the current distribution. Both
citizens learn each other's opinion and each of them compares the
distance in 
opinions, i.e. the absolute value of opinion difference, with her
individual threshold $\sigma _{i}^{t}$, also represented by a continuous
variable ranging between $0$ and $1$. If the distance in opinions is lower
than the threshold, the citizen's next--period opinion gets closer to that
of her partner in this match. Otherwise, her opinion remains unchanged in
the next time step. Thus, for a match involving citizens $i$ and $j$, we have:
\begin{equation}
x_{i}^{t+1}=\left\{ 
\begin{array}{c}
x_{i}^{t}+\mu (x_{j}^{t}-x_{i}^{t})\text{ \ if \ }\left\vert
x_{j}^{t}-x_{i}^{t}\right\vert \leq \sigma _{i}^{t} \\ 
x_{i}^{t}\text{ \ \ \ \ \ \ \ \ \ \ \ \ \ \ \ \ \ \ if \ }\left\vert
x_{j}^{t}-x_{i}^{t}\right\vert >\sigma _{i}^{t}
\end{array}
\right.  \label{rule}
\end{equation}
and similarly for $x_{j}^{t+1}$. Given that the convergence parameter $\mu $
does not fundamentally affect the results (see Neau~\cite{Neau}), we set
$\mu=0.5$ to minimize computation time in simulations.  

Note that the threshold $\sigma _{i}^{t}$ is individual and dynamic. This is
different from the basic model of Deffuant et al.~\cite{Deffuant}, where all
citizens have the same constant threshold, as well as from the extended
model of Weisbuch et al.~\cite{Weisbuch}, where the threshold can change in
relation to the variance of the distribution of opinions sampled by the agent.

We assume that the following rule links an agent's opinion to her threshold: 
\begin{equation}
\sigma _{i}^{t}=\alpha \Delta _{i}^{t}-\gamma (\Delta _{i}^{t})^{2}\, ,
\label{sigma}
\end{equation}
where $\Delta _{i}^{t}=\max \{x_{i}^{t},1-x_{i}^{t}\}$
measures the distance of the opinion of agent $i$ at time $t$ from the
furthest extreme opinion. The individual threshold is thus shaped
by two forces:
\begin{itemize}
\item[(i)]{\bf Benefits of diversification}, $\alpha \Delta _{i}^{t}$. It
  captures the 
hypothesis that a citizen closer to one extreme (thus further from the other
extreme) is likely to meet someone whose opinion is very
different from hers, and thus can potentially reap large benefits of
diversity. For instance, people at the
two extremes may get information from very different sources and their
social interaction can help both of them to get a highly informative a posteriori signal
about the true state of the world. Thus, we assume that the opinion distance from the other
extreme has a positive linear relationship with the threshold.

\item[(ii)]{\bf Communication costs}, $\gamma (\Delta _{i}^{t})^{2}$. The
  likelihood of 
meeting someone with a very different opinion implies
that the communication costs with such a person are also likely to be
higher. Moreover, higher distance implies lower trust, as shown
empirically by Alesina and La Ferrara~\cite{Alesina}. It is plausible that the increase in the communication cost (as the
distance to one extreme decreases) is not constant but growing. We
capture this second channel by a negative quadratic relationship between the
distance from the furthest extreme and the threshold.
\end{itemize}

We do not know a priori which of the two channels dominates and shapes opinion
dynamics. We therefore study the behavior of the social system for a large range of parameter values $\alpha $ (that describes the incremental
benefit of diversity) and $\gamma $ (describing the increase in the
incremental cost of communication). Since we are interested in the
\textit{relative} weight of these two channels, we introduce an additional
parameter, $\beta =\frac{\gamma }{\alpha }$, which allows us to
rewrite~\eqref{sigma} as follows:
\begin{equation}
\sigma _{i}^{t}=\alpha \Delta _{i}^{t}\left(1-\beta \Delta _{i}^{t}\right)\, .
\label{sigma2}
\end{equation}

These two channels have also a sociological
interpretation. $\alpha $ captures the 
motivation that makes people to construct friendship relations: the more
friends you have the easier you can find help when in trouble. $\gamma $
represents the dissonance force that prevents one from becoming a
friend with someone that holds a too distant opinion.

Typically, after a certain number of encounters the opinion distribution starts
to exhibit one or more clusters. These are sizeable groups of citizens holding
relatively close opinions. Eventually, the social system reaches an asymptotic state such
that the opinion distribution becomes stationary.

This asymptotic state will be mainly characterized by the following two
measures: the number of clusters that form and the cluster formation 
time. We define these measures more precisely below. We also characterize and study the
existence and the number of large and small clusters.

\begin{center}
 \begin{figure}[ht]
  \begin{center}
  \makebox{\includegraphics[scale=0.18,angle=0]{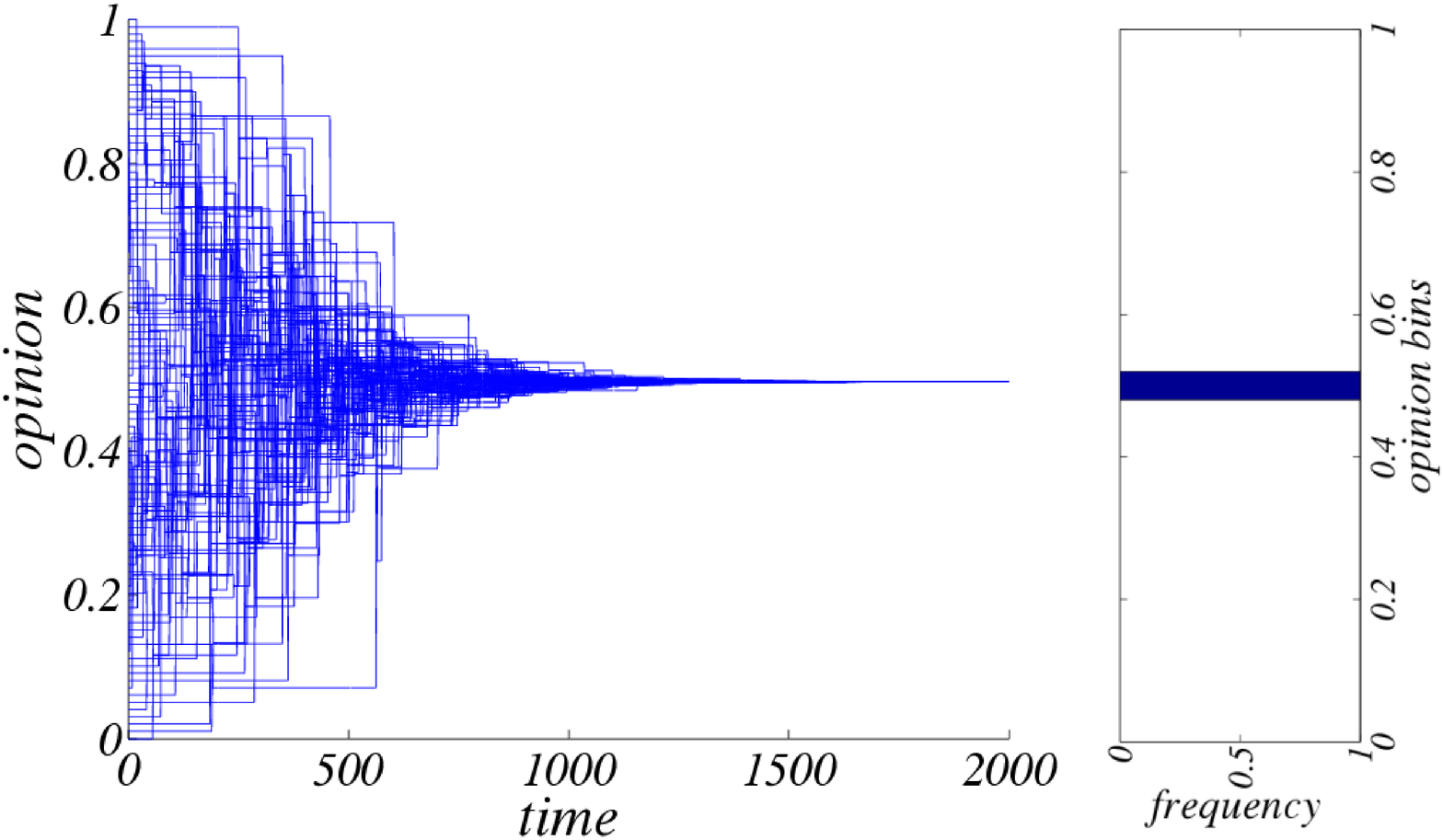}}\quad
  \makebox{\includegraphics[scale=0.18,angle=0]{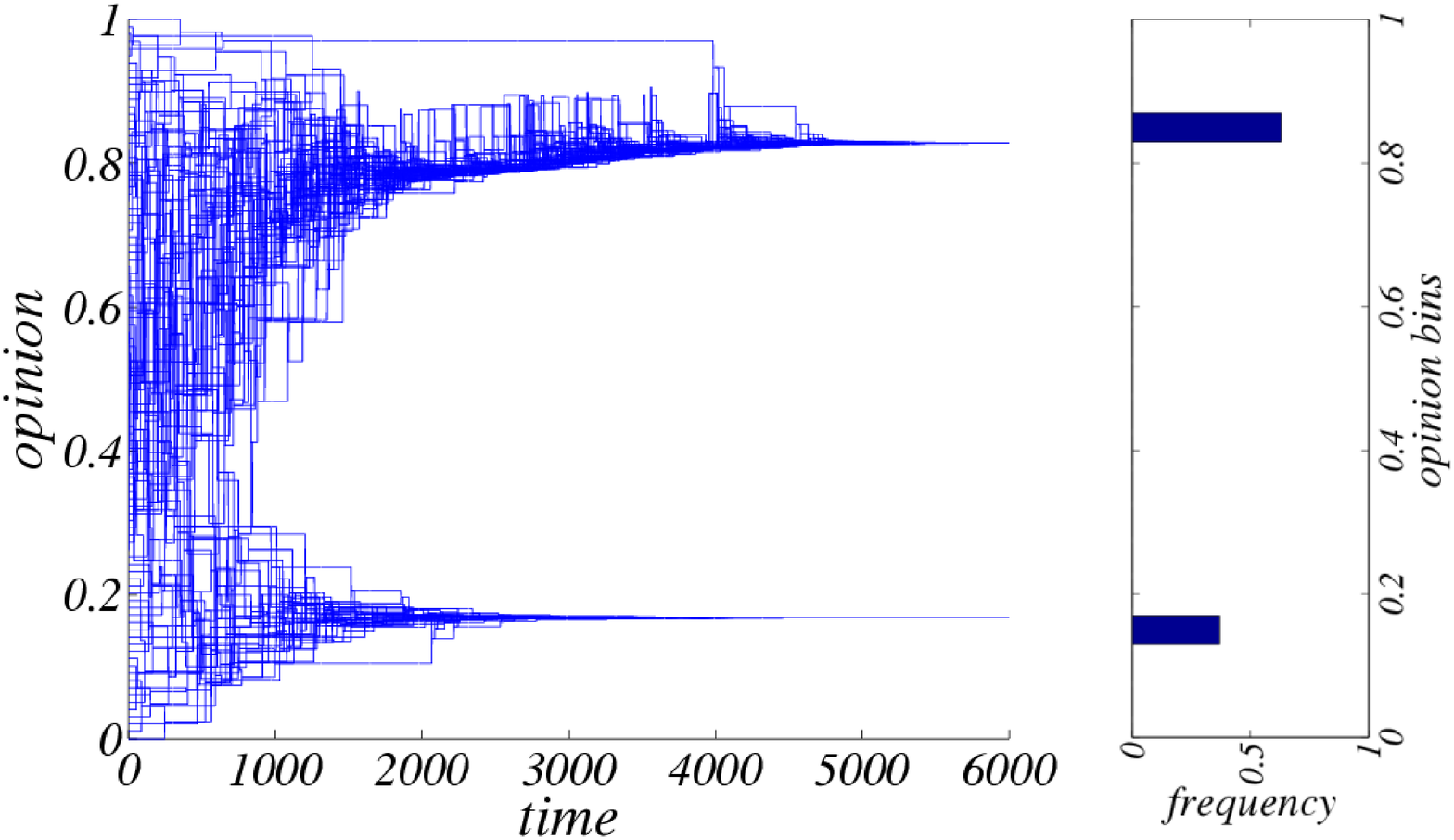}}
  \caption{Two typical realizations. Both refer to $N=100$
    individuals with initially uniformly distributed opinions in $[0,1]$. In
    the left panel (a single cluster case) parameters are $\alpha = 1.0$ and
    $\beta =0.6$; in the right panel (two clusters case), $\alpha = 1.0$
    and $\beta =0.9$. On the right of each panels -- the final distribution
  of opinions representing the number of clusters.}
  \label{fig:figure1}
  \end{center}
 \end{figure}
 \end{center}

Figure~\ref{fig:figure1} presents two typical realizations of the numerical
simulations of 
our model. For both realizations we assume $N=100$ agents with an initial
distribution of 
opinions being uniform on $[0,1]$ interval. The realization in panel (a)
corresponds to the case with $\alpha =1$ and $\beta =0.6$, and it exhibits
consensus: there is a unique cluster of opinions in the asymptotic
state. The realization in 
panel (b) is obtained under parameters $\alpha
=1$ and $\beta =0.9$ and corresponds to the polarization case. 

The first step in our analysis is to define a cluster more precisely. Because of the dynamics of the model, in the 
asymptotic state all clusters are clearly separated. We divide the whole opinion space in uniform bins (we used bins with width
$5/N$) and we define a cluster as a group of agents whose opinions are
contained in neighboring bins. Moreover a cluster is \textit{large} if
it contains more than $5\%$ of the total number of agents; otherwise we call
it a \textit{small} cluster.

In the asymptotic state, each cluster has an associated cluster
opinion, i.e. the opinion shared by all agents in the cluster. An
agent is aggregated into the cluster if her opinion differs from the cluster
opinion by less than a certain cutoff (we use the cutoff equal to
$0.0001$). This is an \textit{a posteriori} measure which we need 
to pin down the asymptotic state. \ 

Finally, we define the cluster formation time, $T_{c}$, as the number of time steps
needed to aggregate into the cluster at least $95\%$ of the total 
number of agents that the cluster will contain in the asymptotic state.

\section{Analysis of the number of clusters}

The first characteristic of interest is the number of (large and small)
clusters of opinion in the asymptotic state. From the applied point of view,
this characteristic is of key importance, because it describes whether the
public opinion in the population will tend to a \textit{consensus} (a single
large cluster), \textit{polarization} (two large clusters), or \textit{pluralism} (more than two clusters). One can plausibly hypothesize that it
is exactly this characteristic that determines whether the community will
adopt a political system with two moderate parties that have convergent
political platforms (as, for example, in Australia), two parties with
relatively distinct platforms (as in the United States and the UK), or a
multiple party system with resulting coalition governments (as in Italy and
Netherlands).

\begin{center}
 \begin{figure}[ht]
  \begin{center}
  \makebox{\includegraphics[scale=0.4,angle=0]{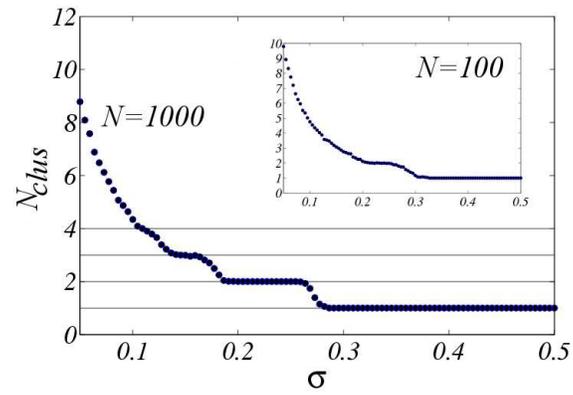}}
  \caption{$1/(2\sigma)$--rule. We present here the results of two simulations
    for the determination of the number of large clusters as a function of the threshold $\sigma$ in the Deffuant et
    al. model~\cite{Deffuant} with $N=1000$ 
    agents. The inset shows the results for $N=100$ agents.}
  \label{fig:figure2}
  \end{center}
 \end{figure}
 \end{center}

In the basic continuous--opinion dynamics model with a unique constant
threshold~\cite{Deffuant,Weisbuch}, the authors establish the so--called
$1/(2\sigma )$--rule for the average 
number of large clusters of opinion in the asymptotic state. This rule states that
the threshold $\sigma $ completely determines the average number of large clusters,
which is equal to the integer part of $1/(2\sigma )$. This is
presented in Figure~\ref{fig:figure2} for a population of $N=1000$ agents; we
also report a simulation with $N=100$ citizens (see inset of
Figure~\ref{fig:figure2}), for comparison. Despite the good
description of the asymptotic state given by that rule, there are nevertheless
some discrepancies due to the finiteness of the population size.

We expect our model to behave differently. In particular, the
average number of large clusters, denoted by $N_{clus}$, and the
average number of small clusters, $n_{clus}$, should depend on
both parameters $\alpha$ and $\beta$. The individual--level
benefit--of--diversity parameter $\alpha $ represents in the aggregate the
centripetal force of the social system. On the contrary, the individual--level
incremental communication cost parameter $\beta $ represents in the
aggregate the relative strength (with respect to $\alpha $) of the
centrifugal force of the system. It is the combination of these two forces
that eventually determine the dynamics and the asymptotics of the system.

\begin{center}
 \begin{figure}[ht]
  \begin{center}
  \makebox{\includegraphics[scale=0.32,angle=0]{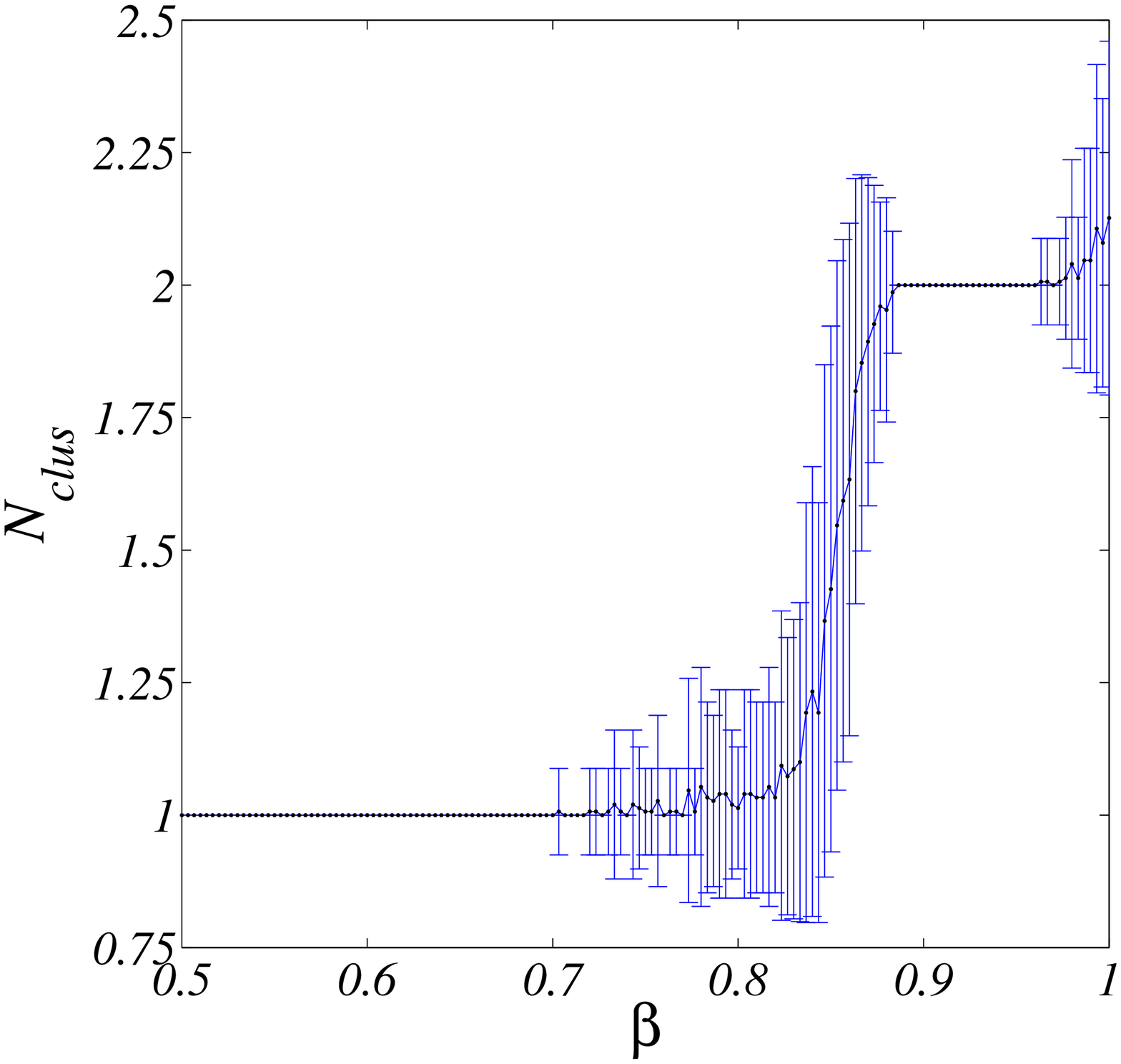}}\quad 
  \makebox{\includegraphics[scale=0.32,angle=0]{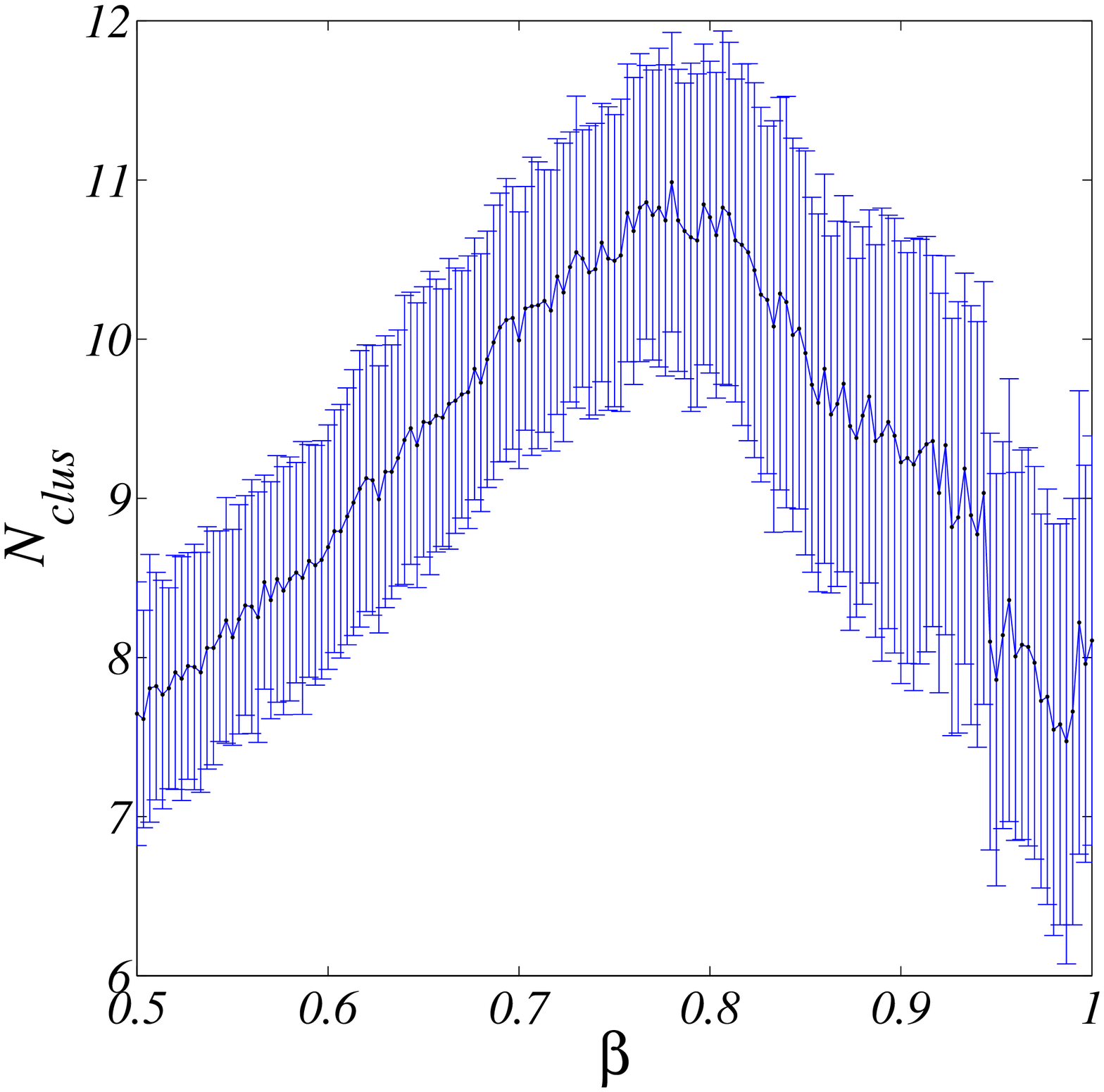}}
  \caption{Average number of large clusters. We present two typical
    realizations of our model showing the dependence of $N_{clus}$ on $\beta$
    for a fixed value of $\alpha$. Left panel: $\alpha=1.2$, right panel:
    $\alpha=0.14$. The whisker bars around the averages represent the standard
    deviation.} 
  \label{fig:figure3}
  \end{center}
 \end{figure}
 \end{center}

Figure~\ref{fig:figure3} shows the evolution of the average number of large
clusters for relatively high fixed values of $\alpha$ (left panel) and low
values of
$\alpha$ (right panel), while $\beta$ varies in $[0.5,1]$. The reported
simulations have been obtained with values $\alpha =1.2$ (left panel) and
$\alpha=0.14$ (right panel), and they do not
qualitatively change for moderate changes in $\alpha$ (results not reported
here). The whisker bars 
around the average represent the standard 
deviation in the number of clusters.

The data presented in Figure~\ref{fig:figure3} allow us to obtain the
following results.

\textit{Result 1}. The relatively high values of $\alpha $ (i.e., high
benefits of diversity) induce a low number of large clusters ($1$ or $2$,
and rarely above $2$) in the whole range of $\beta$. The strong
centripetal force leads either to 
consensus or to polarization, but never to pluralism. We thus obtain a
somewhat paradoxical result that societies where citizens have a strong
tendency to be open to the opinion of others never obtain pluralism of
opinions in the long run. We provide an explanation to this paradox in the
light of our model later in the paper.

\textit{Result 2}. The transition from consensus (one large cluster) to
polarization (two large clusters) is driven (at fixed $\alpha$) by the
increase in the relative 
strength of the centrifugal force $\beta $ (i.e., the incremental
communication costs). In other words, societies where citizens perceive
strong benefits of diversity tend to reach consensus if the incremental
communication costs between citizens with diverse opinions are relatively
low. However, the same societies become polarized if the incremental
communication costs are relatively high.

\textit{Result 3}. In the ($\alpha ,\beta $) parameter range that correspond
to the transition from consensus to polarization, we observe a much higher
standard deviation in the number of clusters. This implies that in
such transition ranges, small random differences in dynamics of opinion
formation at the early periods can lead to large differences in the asymptotic
state. In other words, in such ranges the social system is non--ergodic. From
the applied point of view this means that we can observe societies which are 
\textit{very similar} in their social characteristics (same benefits of
diversity and same communication costs) that nevertheless exhibit totally
different long--run public opinion distributions (consensus in some
societies, polarization in others).

Analyzing the relatively low values of $\alpha$ (the right panel of
Figure~\ref{fig:figure3}) we can state two further results.

\textit{Result 4}. The average number of large clusters in this case is
much higher than in the left panel of the same
figure, varying between $6$ and $12$ for $\alpha=0.14$. This is a complementary
result to the one described above (\textit{Result 1}): the societies exhibit pluralism in public
opinion only when the benefits of diversity are relatively low.

We are able to explain these paradoxes (\textit{Results 1} and \textit{4}) as
follows: for the pluralism of opinion to be a \textit{stable} feature of the
society, the citizens should have relatively little incentives to change
their opinions, i.e. they should be sufficiently conservative. Only this
way an initially pluralist distribution of opinions can remain pluralist in
the long run. Otherwise (that is, if the benefits of diversity are
sufficiently high, as in the left panel of Figure~\ref{fig:figure3}), citizens
will tend to 
adjust their 
opinions towards those of citizens whom they meet, and a sufficiently high
number of random encounters guarantees that eventually the loci of pluralism
disappear, giving way to clustering of all citizens around one or two
opinions.

Note also that for low values of $\alpha$, the average number of large
clusters is not monotonic in $\beta$. In other words, increasing the
relative strength of the centrifugal forces of the social system leads to an
increase in the average number of large clusters only up to some point.
Beyond this point, further increases in the centrifugal forces tend actually
to decrease the average number of large clusters. This is an artificial
outcome related to the definition of large and small clusters. As shown in
the inset of the right panel of Figure~\ref{fig:figure3} and as we will see in
our discussion 
of small clusters below, beyond a sufficiently high number of large
clusters, the distinction between large and small clusters becomes weak.

\textit{Result 5}. The variation in the number of large clusters is much
higher for lower values of $\alpha $. Moreover, for low values of $\alpha $,
the number of large clusters varies a lot throughout the different values of 
$\beta $, while for the high values of $\alpha $ this variation occurs only
in the transition ranges (see the left panel of Figure~\ref{fig:figure3} and
\textit{Result 3} above). Given  
\textit{Result 4}, this means that the pluralism in public opinion is
fundamentally non--ergodic. In other words, it is basically impossible to
predict the exact number of large opinion clusters in pluralist societies,
based on their observable characteristics (benefits of diversity and
communication costs).

We next turn to the analysis of small clusters of opinion. While for a
social system with high number of large clusters the distinction between
large and small clusters becomes weak, for a system with consensus or
polarization, the small
clusters have a precise meaning: they represent small groups of
citizens that do not get aggregated in the mainstream opinion group(s). From the applied point of view, small clusters can represent
marginal minority--opinion groups (as, for example, the most radical
left--wing or right--wing factions).

\begin{center}
 \begin{figure}[ht]
  \begin{center}
  \makebox{\includegraphics[scale=0.32,angle=0]{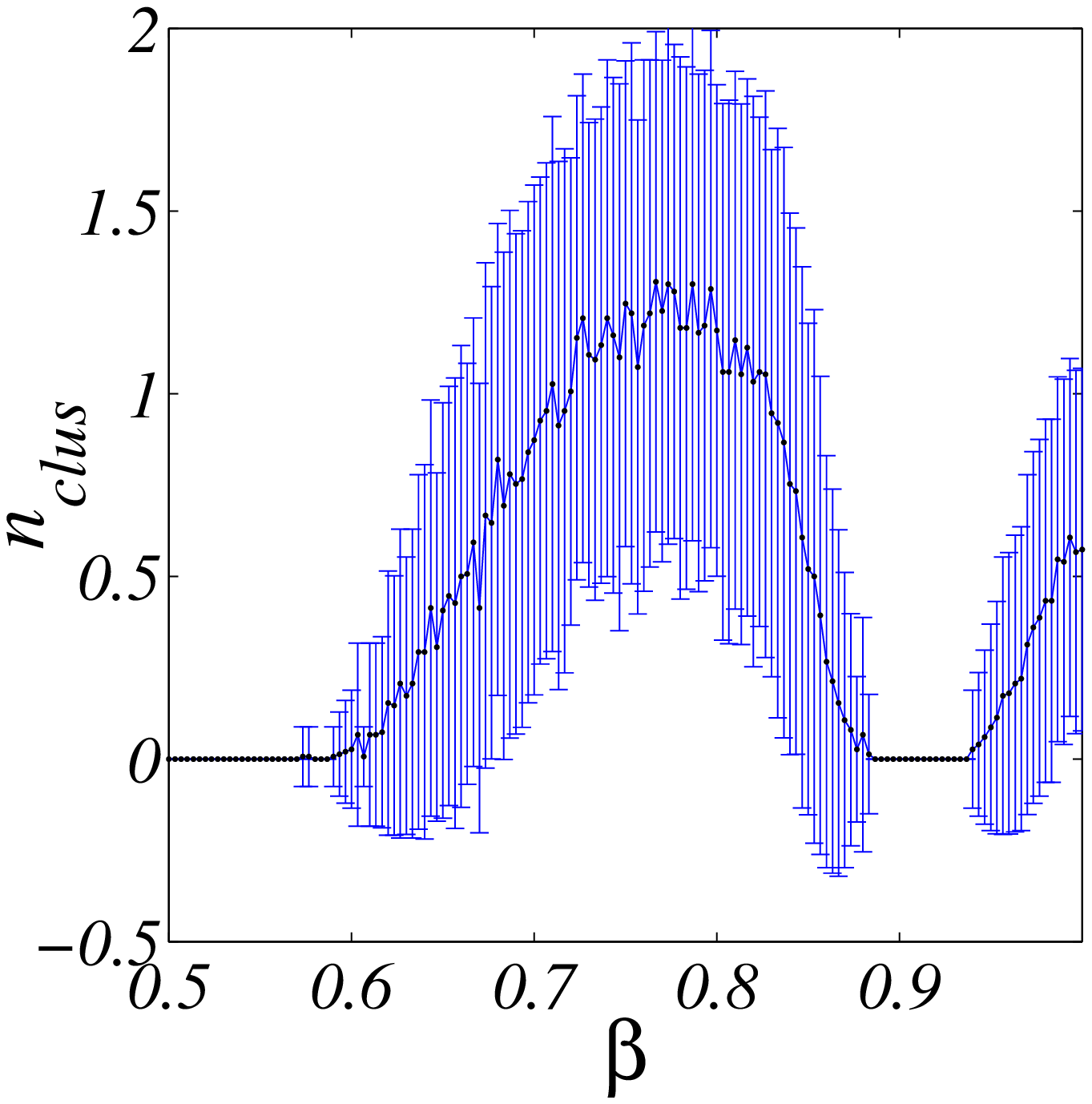}}\quad 
  \makebox{\includegraphics[scale=0.32,angle=0]{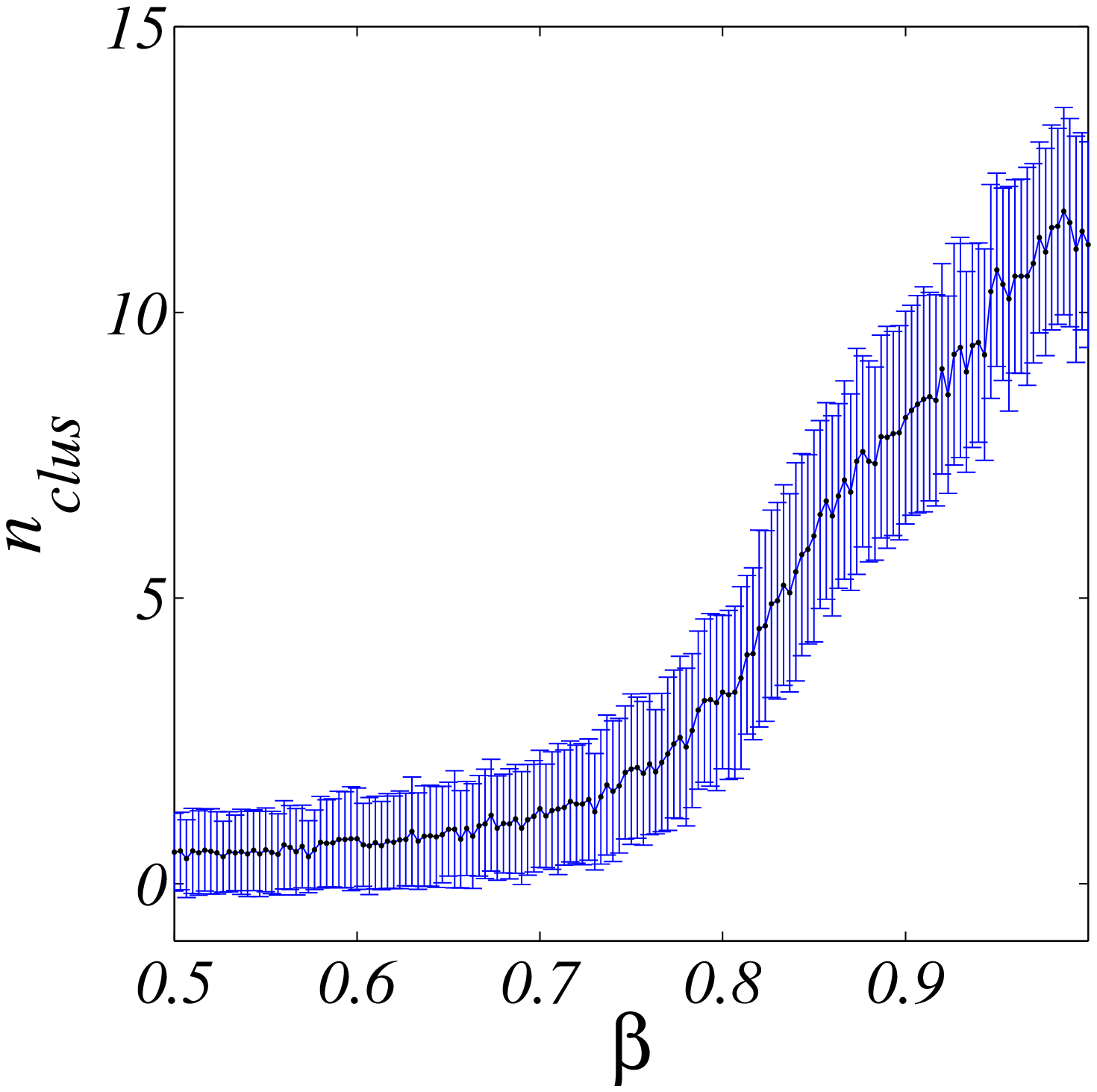}}
  \caption{Average number of small clusters. We present two typical
    realizations of our model showing the dependence of $n_{clus}$ on $\beta$
    for a fixed value of $\alpha$. Left panel: $\alpha=1.2$, right panel:
    $\alpha=0.14$. The whisker bars around the averages represent the standard
    deviation.} 
  \label{fig:figure5}
  \end{center}
 \end{figure}
 \end{center}

This analysis is summarized in Figure~\ref{fig:figure5} where we present the
evolution of the average number of small 
clusters (as a function of $\beta $) for different values of
$\alpha$. The left panel shows the results obtained with $\alpha=1.2$, while the right
panel -- for $\alpha =0.14$.

\textit{Result 6}. In case of high values of $\alpha $ (left panel), small
clusters of 
opinion are absent for most parameter ranges, and appear only in the
parameter ranges corresponding to the transition from consensus to
polarization. The puzzling 
feature is not the appearance of small clusters per se, but their
disappearance beyond the transition range (see the left panel of
Figure~\ref{fig:figure5} for $\beta\sim 0.9$). Moreover, small clusters of
opinion are regularly present in simulations with low values of $\alpha $.
This indicates that the appearance of small clusters of opinion is not
mainly driven by the fundamental characteristics of the social system (i.e.
the values of $\alpha $ and $\beta $), but is inherently linked to the
non--ergodicity of the system.

As discussed earlier, the distinction between large and small clusters
becomes weaker as the number of large clusters grows.
This can be seen in the right panel of Figure~\ref{fig:figure5}: the number of
small 
clusters sharply increases when $\beta $ is close to $1$, which is exactly
the range corresponding to the decline in the number of large clusters in the
right panel of Figure~\ref{fig:figure3}. Large and small clusters
become very 
similar in size, differing by at most a few citizens on average, and their
\textit{total} number is monotonic in $\beta $, as can be seen in the
right panels of Figures~\ref{fig:figure3} and~\ref{fig:figure5}.

\begin{center}
 \begin{figure}[ht]
  \begin{center}
  \makebox{\includegraphics[scale=0.4,angle=0]{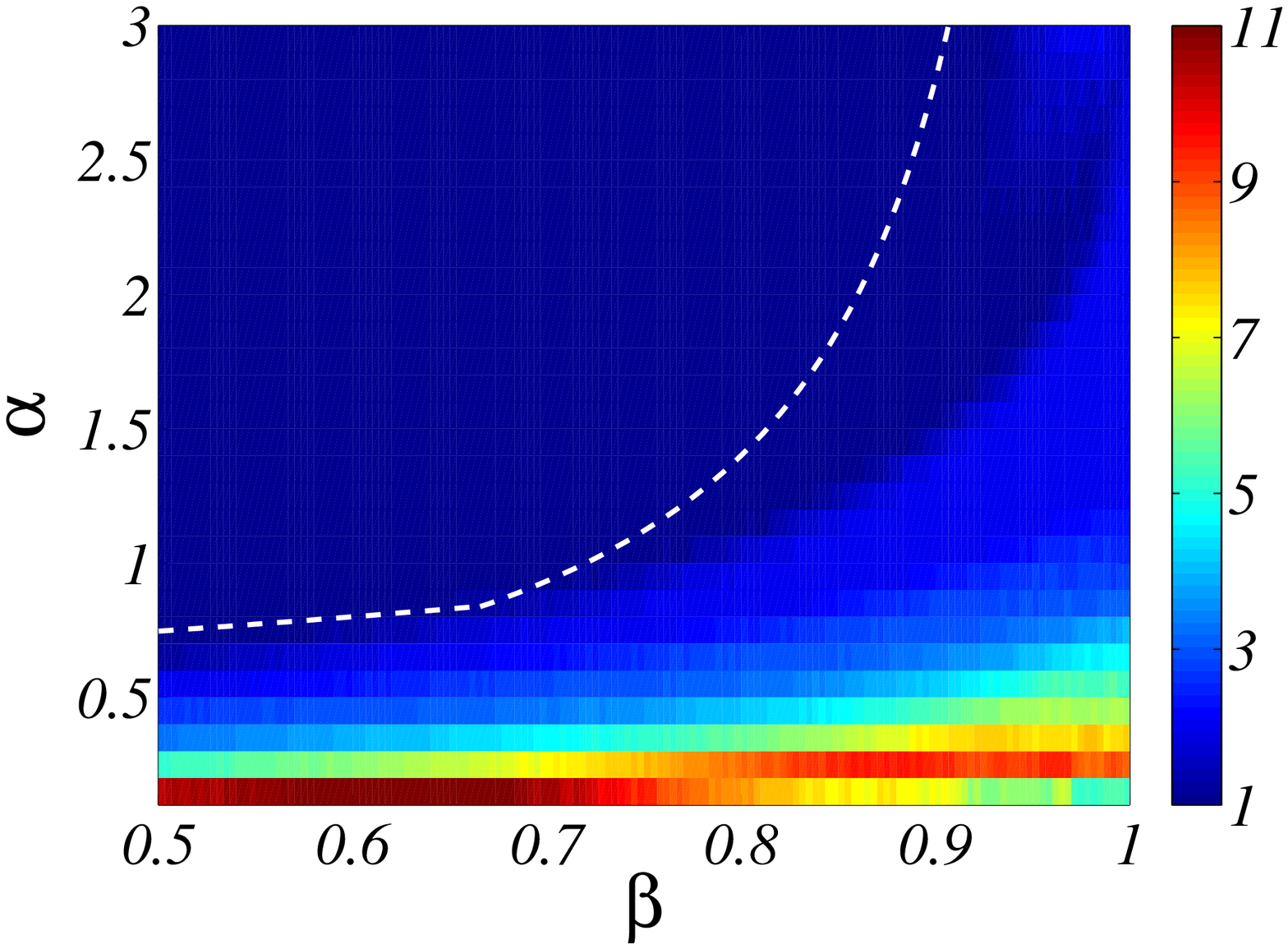}}\quad 
  \makebox{\includegraphics[scale=0.3,angle=0]{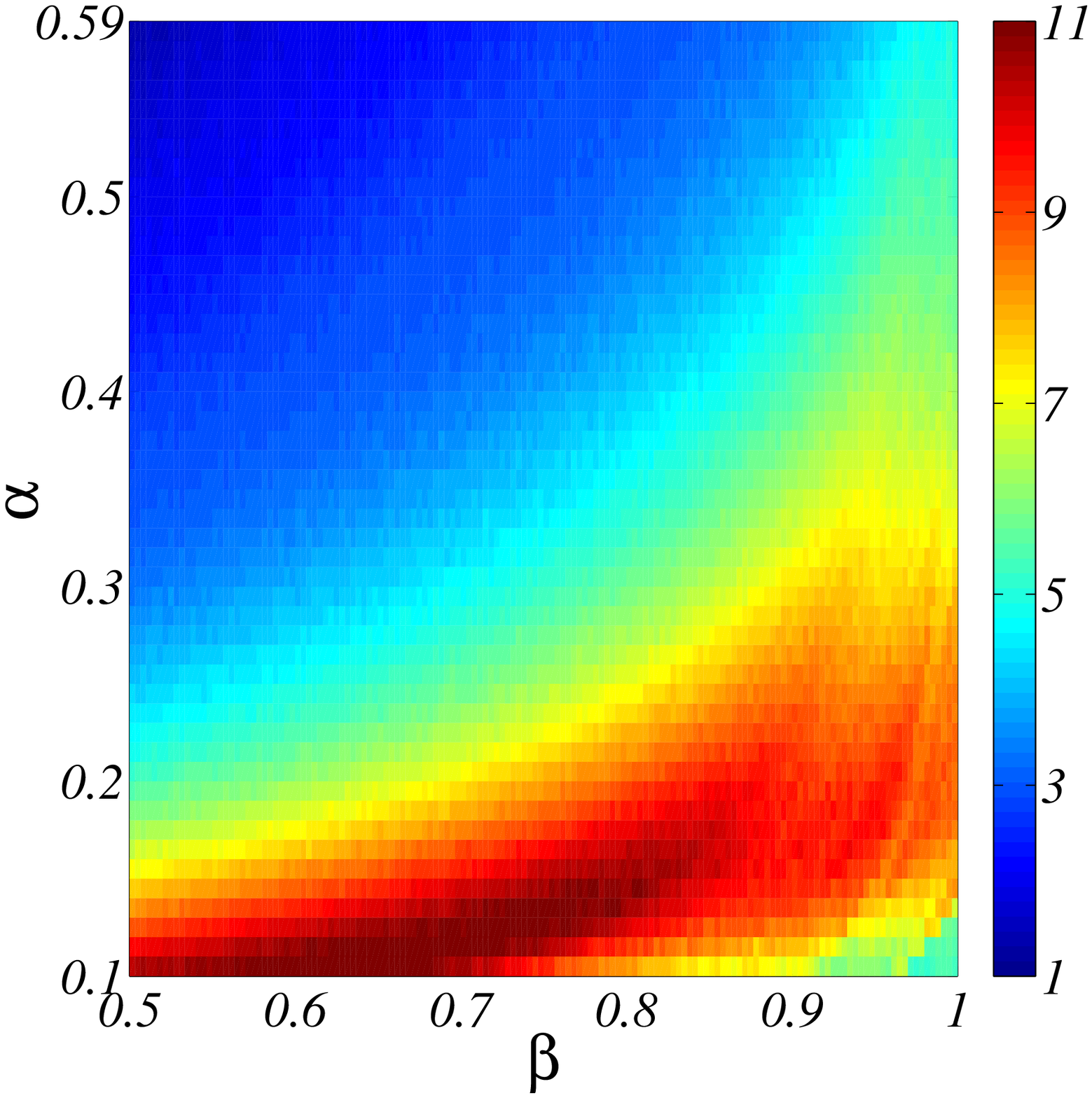}}
  \caption{Number of large clusters as a function of $\alpha$ and
    $\beta$. We present here a global view of the number of large clusters as a function of $\alpha$ and $\beta$ (left panel). The right panel shows a
    zoom in the region $\alpha \in [0.1,0.59]$. The colors denote the number
    of 
    large clustes. The dashed line denotes the boundary of the set where all
    agents have initial thresholds larger than $0.28$.}
  \label{fig:figure7}
  \end{center}
 \end{figure}
 \end{center}

While Figures~\ref{fig:figure3} and~\ref{fig:figure5} already give a
reasonably good sense of the main 
mechanisms driving the model, we can get a more complete understanding of the
model by studying the number of (large) clusters in function of continuous
change in both parameters ($\alpha $ and $\beta $). This analysis is presented
in Figure~\ref{fig:figure7} where we show the average number of large clusters
as a 
function of both $\alpha $ and $\beta $, for the ranges $\alpha \in \lbrack
0.1,3]$ and $\beta 
\in \lbrack 0.5,1]$ (left panel). The right panel presents the zoom in
the range $\alpha \in \lbrack 0.1,0.59]$. These figures have been obtained by
making a grid of a portion of the parameter plane $(\alpha,\beta)$; then,
for any point of the mesh the average number of large clusters is computed
from $150$ simulations. The final results are plotted in a color scale;
increases in the number of clusters corresponds to the move from dark blue 
color towards dark red.

These numerical results can be interpreted in reference to the benchmark unique constant
threshold model~\cite{Deffuant} where the threshold value at which the
transition from consensus to 
polarization occurs is about $0.28$ (for $N=100$). We thus need to find the
loci of ($\alpha ,\beta $) such that all agents have initial thresholds above
this value. The individual thresholds are updated at each interaction resulting
in opinion change and thus they change over time, but we can ensure the
existence of a single cluster if all $\sigma _{t}^{i}$ are larger than $0.28$

In our model the threshold function is a pair of incomplete
parabolae with negative curvature 
and vanishing at $\Delta _{1}=0$ and $\Delta _{2}=1/\beta$
(see~\eqref{sigma2}). Hence, we must 
require that $\beta \leq 1$ to ensure the threshold being positive in the
entire range of opinions. On the other hand, by construction, the threshold
should be lower than $1$. Using its maximum at $\Delta
_{\max }$=$1/(2\beta )$, this implies $\alpha \leq 4\beta $. Finally, each
threshold is 
larger than the smallest between the threshold of the most extreme citizens, $\sigma_{ext}=\alpha (1-\beta )$, and the threshold of most
moderates citizens, $\sigma _{mod}=\frac{\alpha }{4}(2-\beta )$. Note
that for $\beta \in \lbrack 0,2/3]$ we have $\sigma _{ext}\geq \sigma_{mod}$,
while for $\beta \in (2/3,1]$, $\sigma_{ext}<\sigma_{mod}$.

Thus the required condition on the individual thresholds is obtained
under 
\begin{equation}
\min \{\sigma_{ext},\sigma_{mod}\}\geq 0.28.  \label{min}
\end{equation}
The loci of parameters $(\alpha ,\beta)$ that guarantee this condition are
marked in the left panel of Figure~\ref{fig:figure7} by the dashed light
line. The data reported in this figure fully confirms the results discussed
above. One additional result stands out. 

\textit{Result 7}. The comparison with the benchmark model (the dashed line)
indicates that the introduction of heterogeneous dynamic thresholds changes
the range where consensus is obtained. This can be seen in the left panel of
Figure~\ref{fig:figure7}, by the presence of some dark blue area to the right
of the dashed 
line. In this area, a fraction of citizens have initial thresholds that are
below $0.28$ (in the benchmark model, it is the level at which consensus
breaks down); nevertheless, in our model the consensus survives. However,
the shape of this area is not uniform: in particular, for lower levels of
$\beta $, it basically disappears. This implies that for relatively low 
incremental communication costs, there is little qualitative difference
between the benchmark model and our heterogeneous dynamic threshold model.
The heterogeneity of thresholds starts to make a qualitative difference at
the higher levels of incremental communication costs.

\begin{center}
 \begin{figure}[ht]
  \begin{center}
  \makebox{\includegraphics[scale=0.35,angle=0]{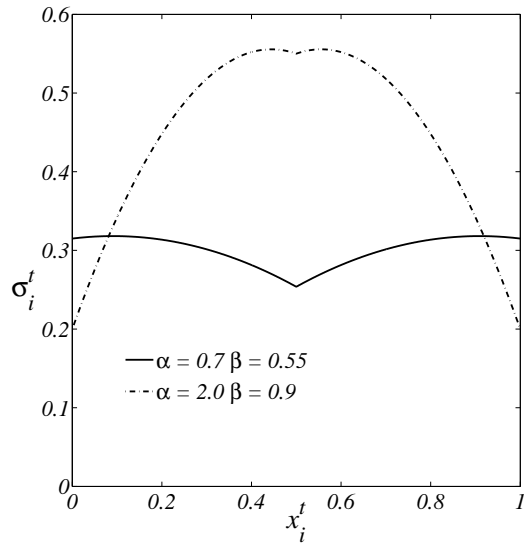}}
  \caption{The threshold $\sigma_i^t$ in function of the opinion $x_i^t$ for
    two sets of parameters $(\alpha,\beta)$.}
  \label{fig:soglia}
  \end{center}
 \end{figure}
 \end{center}

Where does this difference come from? This occurs because for low--$\alpha $/low--$\beta $ combinations slightly above the $0.28$ line, as we move across
the spectrum of opinions, the threshold varies very little. Thus, crossing
the $0.28$ line basically means that we pass from the situation where 
\textit{all} citizens have the threshold above $0.28$ to the situation where 
\textit{almost all }citizens have the threshold below $0.28$, which
basically guarantees that the consensus (a single cluster) is never
obtained. Instead, for high--$\alpha$/high--$\beta$ combinations slightly
above (to the left) the $0.28$ line, as we move across the spectrum of
opinions, the threshold changes non--linearly (sharply around the extremes
and slowly around the center). Thus, crossing the $0.28$ line means that we pass from the situation where \textit{all} citizens have the
threshold above $0.28$ to the situation where almost all citizens still have
the threshold above $0.28$ and \textit{only very few extreme }citizens have
the threshold below $0.28$, which implies almost surely that the consensus
(a single cluster) still obtains (See Figure~\ref{fig:soglia}).

\section{Cluster formation time}

The second characteristic that we are interested in is the cluster formation
time, $T_{c}$. The practical importance of this measure is best explained
through the following example. Consider a community that has to make a
collective decision (via the majority rule) upon some unidimensional issue.
Initially, citizens hold very diverse opinions about the issue; however,
before the vote takes place, some (fixed) amount of time has to elapse,
during which citizen discuss the issue among themselves. A community with
long cluster formation time comes to the vote with a still highly dispersed
distribution of opinions. Therefore, the outcome of the vote will leave
most citizens dissatisfied with the decision. Instead, a community where
cluster(s) form relatively quickly comes to the vote with much more
concentrated opinion distribution (either with one or several clusters).
Thus, the outcome of the vote will satisfy almost all (in case of one
cluster), the majority (in case of two clusters), or a large minority (in
case of more than two clusters) of citizens. In all cases, the latter
community is overall happier with vote outcome, compared to the former
community.

This example indicates the importance of understanding the factors that
determine the cluster formation time. In the benchmark model with a unique
constant threshold, for the single--cluster case, Carletti et
al.~\cite{propaganda} have found that the cluster formation time approximately
obeys 
the following law: 
\begin{equation}
T_{c}\sim \frac{N}{2\sigma }\log \left( 2\sigma N\right) \,,
\label{eq:cottime}
\end{equation}
which (roughly speaking) comes from the $1/(2\sigma )$--law and the mean
number of interactions needed to aggregate the opinions into a single
cluster.

In our model with heterogeneous dynamic thresholds the cluster formation
time is normally longer. Note that on this measure it is impossible
to compare our model with the benchmark model as in our model there isn't a
single threshold, but a distribution of thresholds. However, we still can
fix a certain combination ($\alpha ,\beta $) (such that a
single cluster of opinion almost surely obtains) and compare the cluster
formation time with the one of the benchmark model using the threshold of
extremists, $\sigma =\alpha (1-\beta )$, that of most moderate agents, $\sigma =\alpha /2(1-\beta /2)$, or the average initial threshold. In
all cases, we obtain a longer cluster formation time for our model. By
measuring the cluster formation time in numerical simulations, and then
inspecting the shape of $T_{c}$, we get the following result.

\begin{center}
 \begin{figure}[ht]
  \begin{center}
  \makebox{\includegraphics[scale=0.35,angle=0]{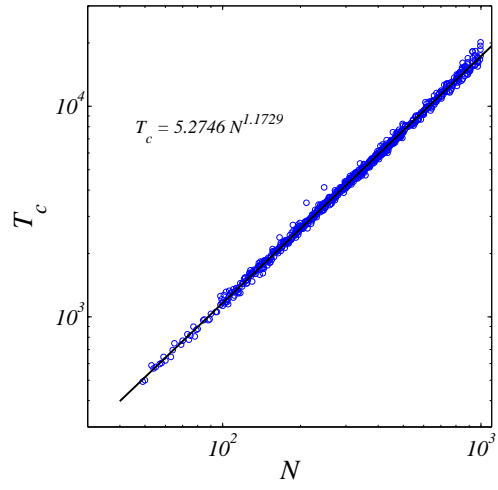}}
  \caption{The cluster formation time as a function of population size
      in the single--cluster case. The straight line in the log--log plot
      represents the fitted relation~\eqref{eq:cotdtime}.} 
  \label{fig:figure8}
  \end{center}
 \end{figure}
 \end{center}

\textit{Result 8}. The cluster formation time approximately follows a power
law in the community size: 
\begin{equation}
T_{c}\sim AN^{B}\,,  \label{eq:cotdtime}
\end{equation}
where the constants $A$ and $B$ depend on the parameters $\alpha $ and $\beta
$. Some values of these constants are reported in Table~\ref{tab:table1} for
different parameters $(\alpha,\beta)$. These numerical values have been
obtained performing a linear regression on the data in the log--log variables,
presented in Figure~\ref{fig:figure8} in the
single cluster case (parameters $\alpha =1$ and $\beta =0.6$), where we show
the dependence of $T_c$ on the total 
number of agents in a log--log plot.

Clearly, given that in each period there is only one encounter, larger
communities take more time to aggregate into the single cluster. Moreover,
by the shape of the power law, the incremental increase in the cluster
formation time is stronger for larger communities. However, the incremental
effect of the community size on the speed of aggregation crucially depends
on the parameters $\alpha $ and $\beta $. This can be seen in
Table~\ref{tab:table1}: 
varying $\alpha $ and $\beta $ leads to changes in the coefficients $A$ and
$B$. This can be summarized in to the following result.

\textit{Result 9}. Both an increase in $\beta $ for a given $\alpha $ and a
decrease in $\alpha $ for a given $\beta $ lead to reinforce the incremental
effect of the community size on the speed of aggregation. In other words,
lower thresholds act much stronger as barriers to convergence for larger
communities than for smaller ones. The inverse is also true: a small
increase in thresholds (for instance, because of lower incremental
communication costs between citizens) induces a huge fall in the cluster
formation time for a community with 1000 citizens, while the same increase induces a much smaller fall in the cluster formation time for a community
with 200 citizens.

In the case of communities where several clusters of opinion form, the
cluster formation time depends on the number of clusters but also on their
relative 
weights. For a given number of agents in the cluster, the formation time is
longer if more clusters are present. This is because of the existence of a
transient phase where citizens adjust their opinions quite often at the
early stage (See the right panel of Figure~\ref{fig:figure1}). The problem of
cluster formation time where more than one 
cluster are present has a deep impact on the dynamics of the social
system. To the best of our knowledge, even in the benchmark model no theory is
available regarding cluster formation time for the multi--cluster asymptotic
states. Here we establish some preliminary results in this direction.

To do so, we run $500$ simulations for communities with size that varies
from $100$ to $500$ individuals by steps of $10$. We fix parameter values to
ensure the existence of two clusters on average in the asymptotic state;
results used to compute Table~\ref{tab:table2} and the relation~\eqref{regfit}
have been obtained using $\alpha =1$ and $\beta =0.6$. From the
sample created in this manner, we eliminate the cases with more than two
clusters (since in some cases small clusters are formed). For all the remaining
cases, we analyze how the formation time of one cluster depends on its size
and the formation time of the other cluster. Given the duality, and treating
each of the two clusters in a single simulation independently, we can use
the results of each simulation twice. This gives us the sample of 818
observations. We thus look at the following measures: the first cluster
formation time ($T_{c}$), its size ($N_{c}$), and the formation time of its
\lq\lq rival\rq\rq cluster ($T_{c}^{r}$).

Next, we run a linear regression (in logs) using the following statistical
specification:
\begin{equation}
\log T_{c}=C+\eta \log N_{c}+\rho \log T_{c}^{r}+\varepsilon \, ,  \label{reg}
\end{equation}
where $\varepsilon $ is the error term, for which we assume the standard
linear regression assumptions to hold. Table~\ref{tab:table2} presents the regression
results.

We observe that our statistical model fits the simulation data well (it
explains more than $72\%$ of variation in the data). All the three
regression coefficients ($C$, $\eta $, and $\rho $) are positive and highly
statistically significant. This allows us to describe the data with the
following fitted model:
\begin{equation}
\log T_{c}=0.978+1.597\log N_{c}+0.100\log T_{c}^{r}\, .  \label{regfit}
\end{equation}

This analysis allows us to formulate a final result.

\textit{Result 10}. The cluster formation time of different clusters (in
the two-cluster case) is positively correlated. In other words,
independently of the community size, a faster formation of one cluster of
opinion induces an increase in the formation time of the second cluster.
Intuitively, the quick formation of one cluster has two opposed effects on
the dynamics of the other cluster: on one hand, a quick separation of
clusters acts in the same way as the reduction in community size (basically
as if we were having two smaller communities with one cluster each). From
the above discussion we know that smaller communities converge faster.
However, on the other hand, the two clusters are still linked through random
encounters, and thus the part of encounters that happens across clusters is
effectively \lq\lq wasted\rq\rq  for cluster formation. This effect works in the
opposite direction. The regression results show that the first effect
dominates. However, given that the size of the coefficient $\rho $ is
relatively small, we can hypothesize that the second effect is also quite
important. 

Naturally, these results crucially depend on parameters $\alpha $ and
$\beta$. We believe that these results are intriguing  but a more complete
analysis is needed, in particular to test the robustness of our estimates to
changes in the fundamental 
parameters and the variation in regression coefficients within the
two--cluster area of Figure~\ref{fig:figure7}. We leave this analysis for
future work. 

\section{Conclusions}

In this paper we have shown that introducing heterogeneous dynamic thresholds
into the 
model of continuous opinion dynamics and grounding those thresholds in the
simple and intuitive characteristics of individual cost--benefit analysis
substantially deepens our understanding of public opinion dynamics and
generates several new results. We understand better why some
societies converge towards consensus, others become polarized, while some
end up having pluralist public opinion. We link these outcomes, on one hand,
to the benefits of diversity and on the other, to the communication costs
between citizens. We also discover new insights regarding the time
needed for clusters of opinions to form, in particular for polarized
societies case.

More generally, this paper shows that a more detailed specification of the elements of agents' behavior at the individual level
opens many interesting avenues also at the aggregate level. We hope
that our analysis and results will motivate the public opinion researchers to
follow this route.

\newpage

\newpage

\newpage

\newpage


\newpage
\newpage

\newpage

\newpage
\begin{table}[hp]
\begin{center}
   \begin{tabular}[h]{c|c|c}
\hline\hline $(\alpha,\beta)$& A & B \\ \hline
$(2,0.6)$ & 1.2086 & 3.5351 \\
$(2,0.7)$ & 1.1995 & 3.836 \\
$(1,0.6)$ & 1.1729 & 5.2746\\
$(1,0.7)$ & 1.1369 & 7.6978
    \\\hline\hline 
   \end{tabular}
\vspace{1em}
\caption{Cluster formation time (consensus). We report here the numerical values of
  the 
  parameters $A$ and $B$ as a function of $(\alpha,\beta)$ in the
  single--cluster case, $T_c\sim A N^B$.}
  \label{tab:table1}
\end{center}
 \end{table}

\newpage
\begin{table}[hp]
\begin{center}
\begin{tabular}[h]{c|c|c}
\hline\hline 
{\textbf{Variable}} & \textbf{Coefficient} & (Std. Err.) \\ \hline
Log of cluster 1 size & 1.597& 0.066 \\
Log of cluster 2 formation time & 0.100& 0.066 \\
Intercept & 0.978& 0.194\\
\hline \end{tabular}\\
Dependent variable : log of cluster $1$ formation time\\ $N = 818$\, ,
\,\, $R^2=0.725$ 
\vspace{1em}
\caption{Cluster formation time (polarization). The regression results
  of the statistical model~\eqref{reg}.}
\label{tab:table2}
\end{center}
\end{table}

\end{doublespace}
\end{document}